\documentclass[nofootinbib,superscriptaddress]{revtex4}
\usepackage{graphicx}
\usepackage{subfigure}
\usepackage{dcolumn}
\usepackage{bm}
\usepackage{amsmath,amsthm,amssymb}

\begin{document}
%


\title{Bifurcation and fine structure phenomena in critical  collapse of a self-gravitating $\sigma$-field}
\author{Peter C. Aichelburg}
\affiliation{\small{Institut f\"ur Theoretische Physik,
Universit\"at Wien, Wien,
 Austria}}
\author{Piotr Bizo\'n}
\affiliation{\small{M. Smoluchowski Insitute of Physics,
Jagiellonian University, Krak\'ow,
 Poland}}
\author{Zbislaw Tabor}
\affiliation{\small{Department of Biophysics, Jagiellonian
University, Krak\'ow, Poland}}

\date{\today}
\begin{abstract}
Building on previous work on the critical behavior in gravitational
collapse of the self-gravitating $SU(2)$ $\sigma$-field and using
high precision numerical methods we uncover a fine structure hidden
in a narrow window of parameter space. We argue that this numerical
finding has a natural explanation within a dynamical system
framework of critical collapse.

\end{abstract}

\maketitle
\subsection*{Introduction}
Over the past few years the Einstein--$SU(2)\!-\!\sigma$ model has
attracted a great deal of attention \cite{w1, bw1, ch, w2, bw2, bsw,
s}. This model is interesting because its rich phenomenology is
sensitive to the value of a dimensionless parameter $\eta$
 which leads to various bifurcation
phenomena. The most interesting bifurcation was found by Lechner et
al. \cite{w2} who showed that the critical behavior in gravitational
collapse changes character from continuous to discrete
self-similarity when the coupling constant $\eta$ increases above a
critical value $\eta_c$. This phenomenon was interpreted in terms of
dynamical systems theory as the homoclinic loop bifurcation where
the two critical solutions, continously self-similar (CSS) and
discretely self-similar (DSS), merge in phase space. Since the
echoing period $\Delta$ of the DSS solution diverges as
$\eta\rightarrow \eta_c$, the numerical analysis of this bifurcation
is extremely difficult and for this reason some of the aspects of
critical behavior near the bifurcation point, in particular the
black hole mass scaling law, were left open in \cite{w2}.

Below, using high precision numerical methods, we confirm the main
findings of \cite{w2}. In addition, we find that just above the
bifurcation point the marginally supercritical side of the
transition between dispersion and black holes exhibits a fine
structure which is due to the competition between two coexisting
critical solutions, the DSS one and the CSS one. The description of
this phenomenon and its interpretation is the  main purpose of this
paper. The rest of the paper is organized as follows. For readers'
convenience, in section~2 we first briefly repeat the basic setting
of the model and then we summarize what is known about it. In
section~3 we present numerical results and finally, in section~4, we
interpret them.
\subsection*{The model}
The spherically symmetric Einstein-$SU(2)-\sigma$ system is
parametrized by three functions: the metric coefficients $A(t,r)$,
$\delta(t,r)$ and the $\sigma$-field $F(t,r)$, which satisfy the
following system of equations
\begin{equation}
\Box_g F - \frac{\sin(2 F)}{r^2} = 0,\quad \Box_g = -e^{\delta}
\partial_t(e^{\delta} A^{-1} \partial_t) +\frac{e^{\delta}}{r^2}
\partial_r(r^2 e^{-\delta} A\:
\partial_r),
\end{equation}
\begin{eqnarray}
{\partial_t A} &=& -2 \eta\: r A (\partial_t F) (\partial_r F),
\\
{\partial_r \delta} &=& -\eta\: r \left((\partial_r F)^2 + A^{-2}
e^{2\delta} (\partial_t F)^2 \right),
\\
\partial_r A &=& \frac{1-A}{r} - \eta\: r \left( A (\partial_r F)^2 + A^{-1}
e^{2\delta} (\partial_t F)^2 + 2 \:\frac{\sin^2{\!F}}{r^2}\right),
\end{eqnarray}
where $\eta$ is a dimensionless coupling constant. For $\eta=0$ this
system reduces to the $\sigma$ model in Minkowski spacetime. The
initial value problem for this system was studied by Bizo\'n et al.
\cite{bct} for $\eta=0$ and by Husa et al. \cite{w1} for $\eta>0$.
In these studies an important role is played by self-similar
solutions. A countable family of continuously  self-similar (CSS)
solutions, herefater denoted by $CSS_n$ ($n=0,1,...$), was shown to
exist for $0\leq\eta<0.5$ in \cite{b1,bw1,bw2}.
These solutions are regular within the past light cone of the
singularity, however they have a spacelike hypersurface of
marginally trapped surfaces, i.e. an apparent horizon outside the
past light cone if $\eta
> \eta_{n}$, where $\eta_n$ is an increasing sequence ($\eta_{0} = 0.08$, $\eta_{1} = 0.152$, etc.).
 Linear stability analysis
shows that the "ground state" $CSS_{0}$ is stable while the
excitations $CSS_{n}$ have exactly $n$ unstable modes.

 Besides the CSS solutions, the system (1-4) has also a discretely self-similar (DSS)
 solution for $\eta\geq \eta_c\approx 0.17$.
 This solution was constructed by  Lechner \cite{ch}
 via a pseudospectral method following the lines of Gundlach
 \cite{g}.

Next, we summarize what is known about the critical behavior in
gravitational collapse in this model. The first numerical studies of
this problem, reported in \cite{w1}, focused on relatively large
coupling constants $\eta>0.2$. In this range a "clean" type II
critical DSS behavior was observed, however the attempts to resolve
critical evolutions for lower values of $\eta$ encountered numerical
difficulties and for $0.18<\eta<0.2$ only an approximate DSS
behavior was observed. Furthermore the echoing period $\Delta$ was
found to increase sharply  as the coupling constant decreases from
0.5 to 0.18. The critical behavior for smaller couplings
$0.1<\eta<0.2$ was studied in \cite{w2} (still smaller couplings are
less interesting because then the model admits naked singularities).
In the range $0.1<\eta<0.14$ a "clean" CSS critical behavior was
observed, thus it became clear that somewhere  in the interval
$0.14<\eta<0.2$ there must be a transition between CSS and DSS
critical solutions. The detailed studies of this transition
\cite{w2} led to a conjecture that there exist a critical value of
the coupling constant $\eta_{c}\approx 0.17$ for which the system
exhibits the homoclinic loop bifurcation, i.e. the CSS saddle merges
with the DSS limit cycle in the phase space. These results left open
the question which of the two solutions in the transition region
acts as the critical solution at the threshold of black hole
formation. In particular, near the bifurcation point the black-hole
mass scaling could not be properly resolved.
\subsection*{Numerical results}
We have solved equations (1)-(4) for marginally critical initial
data fine tuned to the DSS solution for coupling constants close to
the critical value $\eta_c=0.17$. Since the echoing period $\Delta$
increases sharply as the coupling constant tends to its critical
value from above, it becomes more and more difficult to follow the
evolution over a large number of DSS cycles\footnote{ By an
elementary dimensional analysis the number of cycles scales as
$n\sim -(1/\lambda \Delta) \ln|p-p^*|$, where $\Delta$ is the
echoing period and $\lambda$ is the eigenvalue of the growing mode
of the DSS solution.}. We used the fully constrained implicit
evolution scheme based on the Newton-Raphson iteration. In order to
resolve the singular behavior near the origin we used the grid which
is uniform in $\ln(r)$. To get several cycles of the DSS attractor
near the bifurcation point we had to fine tune parameters of initial
data with precision of $70$ digits - this was achieved with the help
of the arprec library \cite{arp}

Actually, it was not our aim to determine $\Delta$ with high
precision, but rather to show that there exists a $\Delta$  to which
the evolution converges. To this end, we determine $\Delta$ as a
function of time (cycles) as the evolution approaches the limit
cycle i.e. the DSS solution.
For a marginally critical
  solution  we plot
the function $F$ versus
  $\ln(r)$ for some  late time $t_1$ and
  superimpose the profile of the first echo at time $t_2$ shifted by
$\ln(r) \rightarrow \ln(r)+\Delta_r$. The time $t_2$ and the radial
echoing period $\Delta_r$ are chosen to minimize the discrepancy
between two profiles. We also define the temporal echoing period
$\Delta_t$ by the formula $t_2=t^*(1-e^{-\Delta_t})+e^{-\Delta_t}
t_1$, where $t^*$ is the accumulation time. Repeating this
calculation for a sequence of pairs $(t_n,t_{n+1})$, we get a
sequence of values $\Delta_r$ and $\Delta_t$.
 Of
course, if the evolution converges to the DSS solution,  both
$\Delta_r$ and $\Delta_t$ should converge to the same constant. In
Fig.~1
 we show the convergence of
$\Delta_{r}$ and $\Delta_{t}$ during a critical evolution for the
coupling constant $\eta = 0.1725$. Note that the curve levels off,
thus signaling the closeness of the evolution to the limit cycle. As
the coupling constant decreases, $\Delta$ grows and the approach to
the DSS solution becomes slower.

\begin{figure}[h]
\centering
\includegraphics[width=0.4\textwidth,angle=-90]{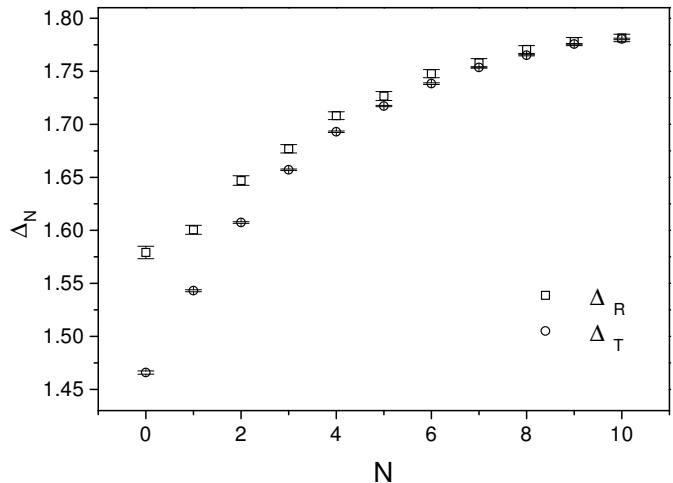}
\caption{\small{For $\eta=0.1725$ we show $\Delta_r$ and $\Delta_t$
as the functions of the cycle number $N$.  Fitting the curve
$\Delta+c e^{-N}$ to the data we obtain $\Delta\approx 1.803$.}
}\label{fig2}
\end{figure}

\begin{figure}[h]
\centering
\includegraphics[width=0.4\textwidth,angle=-90]{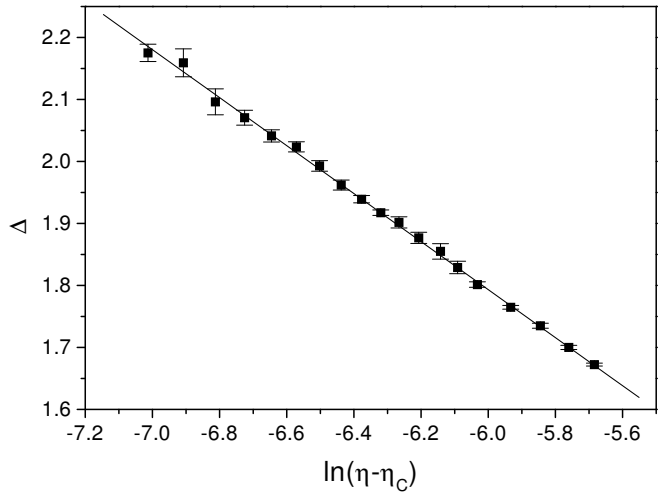}
\caption{\small{Fitting the echoing period $\Delta$ (determined as
in Fig.~1) to the analytic prediction $\Delta = a \ln|\eta-\eta_c| +
const$ we get $\eta_c=0.1701$ and the  slope $a=-0.389$ which is in
very good agreement with the analytic prediction
$a=-2/\lambda_{CSS}$ and  the linear perturbation result \cite{ch}
$\lambda_{CSS}(\eta_c)\approx 5.14$. } }\label{fig2}
\end{figure}

Let $p^*$ be a  critical parameter value which separates dispersion
from black holes (this value can be found by standard bisection). In
agreement with \cite{w1} we find that for $\eta>0.17$ the solution
corresponding to $p^*$ is DSS, in particular for $p$ slightly below
$p^*$ we observe DSS subcriticality, i.e. the solution approaches
the DSS solution and then disperses. Looking at the maximum value of
the spatial derivative of the scalar field at the origin as the
function of $p$, we find a typical subcritical scaling law (see
Fig.~3)
\begin{equation}\label{subscal}
    \max |\partial_r F(t,0)| \sim (p^*-p)^{-\gamma_{DSS}}.
\end{equation}

\begin{figure}[h]
\centering
\includegraphics[width=0.4\textwidth,angle=-90]{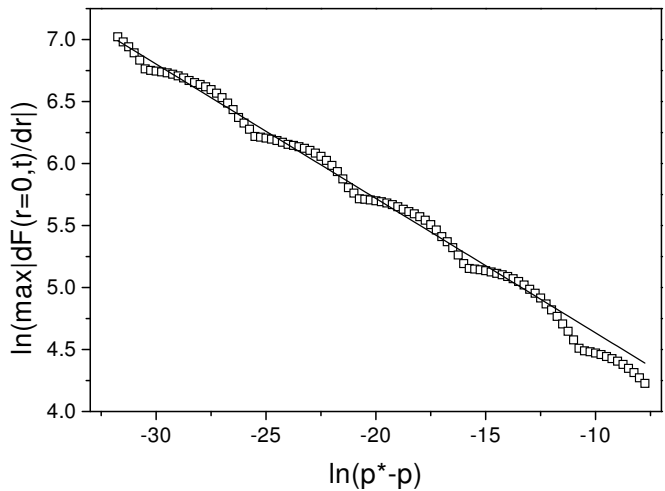}
\caption{\small{The subcritical scaling (5) for the coupling
constant $\eta=0.19$. The slope of the linear fit is approximately
equal to $-0.109$. The wiggles, which are imprints of discrete
self-similarity, have the period $\approx 4.8$.} }\label{fig3}
\end{figure}

For $p > p^*$ black holes are formed, however this happens in a
rather unusual manner. This is shown in Fig.~4 where the metric
function $A$ is seen to develop two minima very close to zero which
signals an almost simultaneous formation of a small and a large
black hole.

\begin{figure}[h]
\centering
\includegraphics[width=0.4\textwidth,angle=-90]{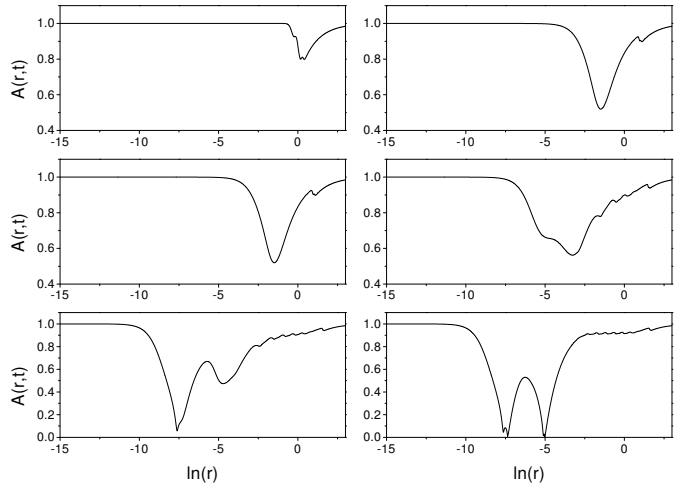}
\caption{\small{The series of snapshots of the metric function
$A(t,r)$ from the evolution of marginally supercritical initial data
for the coupling constant $\eta=0.19$. At the last frame one can see
the formation of two  apparent horizons.} }\label{fig4}
\end{figure}

Let us denote their apparent horizon radii by $r_{in}$ and
$r_{out}$, respectively. We find that the outer radius exhibits the
standard DSS supercritical scaling (see Fig.~5a)
\begin{equation}\label{superscal}
    r_{outer} \sim (p-p^*)^{\gamma_{DSS}},
\end{equation}
but the inner radius does not seem to scale. The latter fact  was
already mentioned in \cite{w2}. The corresponding graph shows a
sea-saw structure, i.e. short straight lines with jump
discontinuities at certain values of the parameter $p$ (see
Fig.~5b).

\begin{figure}[h]
\centering \subfigure[]{
\includegraphics[width=0.34\textwidth,angle=-90]{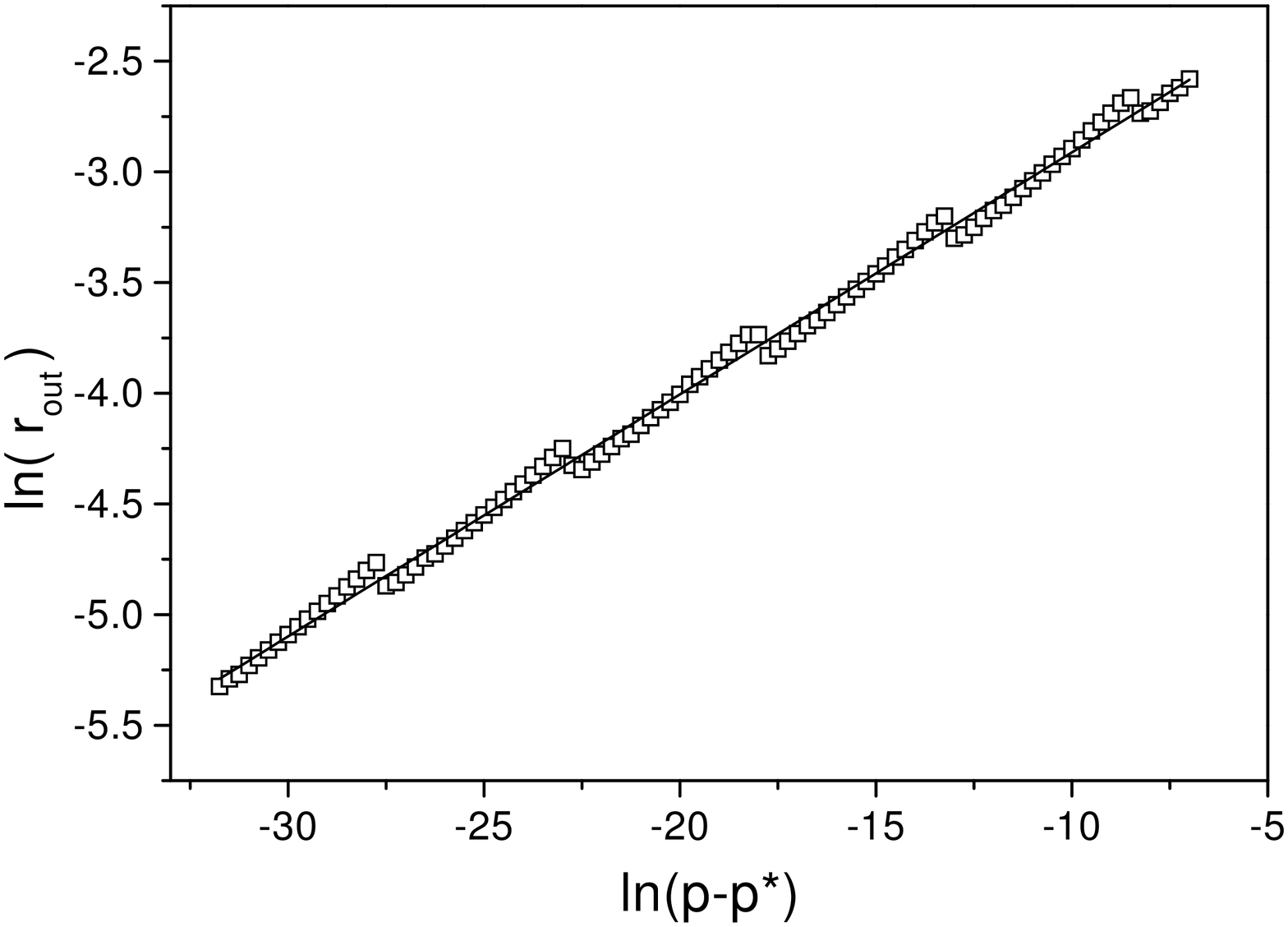}}
\subfigure[]{
\includegraphics[width=0.34\textwidth,angle=-90]{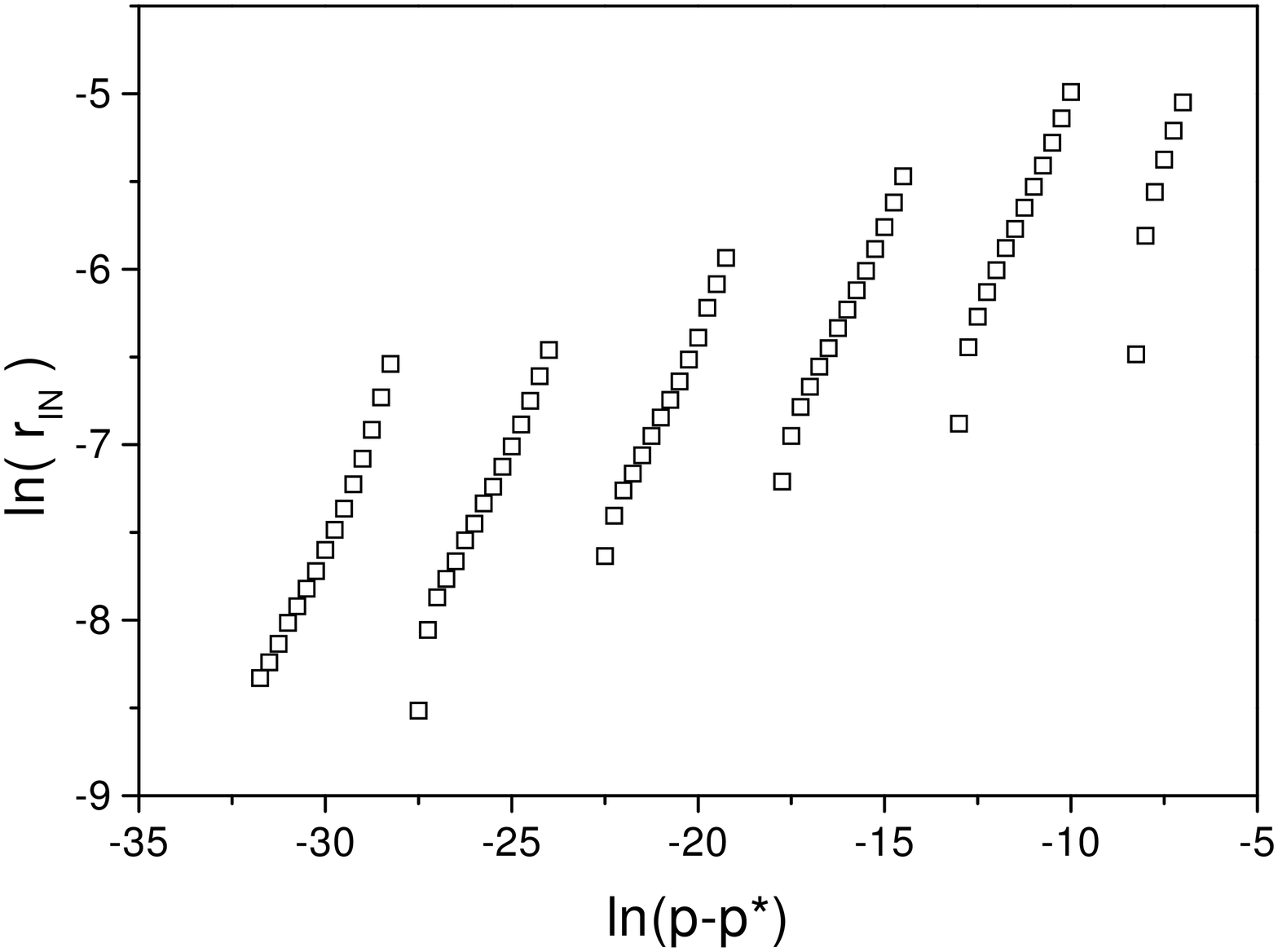}}
\caption{\small{$\eta=0.19$. (a) The locus of the outer apparent
horizon is shown to satisfy the power law (6) with the slope
$\gamma_{DSS}=0.109$. The wiggles superimposed on the linear fit
have the period $4.8$ which agrees with the analytic prediction
$\Delta/2\gamma_{DSS}$. (b) The locus of the inner horizon is shown
not to scale. The jump discontinuities are periodic with the period
$4.8$.} }\label{fig6}
\end{figure}

In order to understand this strange behavior we looked in more
detail at the evolution of initial data fine-tuned to the location
of these jumps.
With the help of high resolution numerical methods we found the
following  remarkable structure: for a given family of initial data
there is a sequence of discrete parameter values $p_{1}> p_{2}
> p _{3}...> p_{n}$ such that a solution with $p\in(p_n,p_{n+1})$
 approaches to the $CSS_{1}$ solution $n$ times, i.e. the
solution comes close to the $CSS_{1}$ solution, turns away and
returns $n$ times before leading to black hole formation.
 Multiple
approaches to the $CSS_1$ solution were already noticed in \cite{w2}
where they were  called episodic CSS, however the corresponding
fine-structure in the parameter space was not seen there. The
sequence $\{p_n\}$ with $n\leq 5$ is listed in Table~1.

\begin{table} [h]
\begin{center}
\begin{tabular}{l||c|c|c|c|c|c}
\hline \hline
$n$       &   1 &      2 &      3 &      4 & 5 & $\infty$ \\
$p_n$ & $0.529001923689295$ & $0.528771570563995$ & $0.528769577618968$
 & $0.528769560376615$ & $0.528769560227152$ & $0.528769560226138$\\
\hline \hline
\end{tabular}
\caption{The first five critical parameter values $p_n$ for the
coupling constant $\eta=0.19$.} \label{tab:pn}
\end{center}
\end{table}

 Because of numerical limitations we were not able to
resolve higher $p_n$, however the data shown in Table~1 seem to
indicate that the sequence $p_n$ converges to $p^*$ as $n$ tends to
infinity. Actually, we find that the  two consecutive parameters
$p_n$ satisfy the scaling law
\begin{equation}
 \frac{p_{n}-p^*}{p_{n+1}-p^*} \approx
 \exp\left(\frac{\Delta}{2\gamma_{DSS}}\right).
 \end{equation}
  Now we return to the problem of scaling of the inner radius
  $r_{in}$.
For $p = p_{n} + \varepsilon$, i.e. for $p$ just above one of the
$p_{n}$'s we see a clear CSS scaling (see Fig.~6a)
\begin{equation}\label{subscal}
    r_{in} \sim (p-p_n)^{\gamma_{CSS}}.
\end{equation}
For $p = p_{n} - \varepsilon$ the solution displays a kind of
pseudo-dispersion after its last CSS-episode.  This
pseudo-dispersion manifests itself as follows: after leaving the CSS
solution, the maximum of the function $F$ decreases, the inner
minimum of $A$ disappears and later a spike develops which leads to
the formation of an apparent horizon at $r_{outer}$. In this range
of $p$ we see the subcritical CSS scaling (see Fig.~6b)
\begin{equation}\label{subscal2}
    \max |\partial_r F(t,0)| \sim (p_{n}-p)^{-\gamma_{CSS}},
\end{equation}
however the masses of black holes formed in such evolutions are
"large" and do not scale. We remark that  since  the solutions on
both sides of $p_n$ form black holes, the bisection which gives
critical parameter values $p_{n}$   has to be performed in a sense
"by hand".

\begin{figure}[h]
\centering\centering \subfigure[]{
\includegraphics[width=0.34\textwidth,angle=-90]{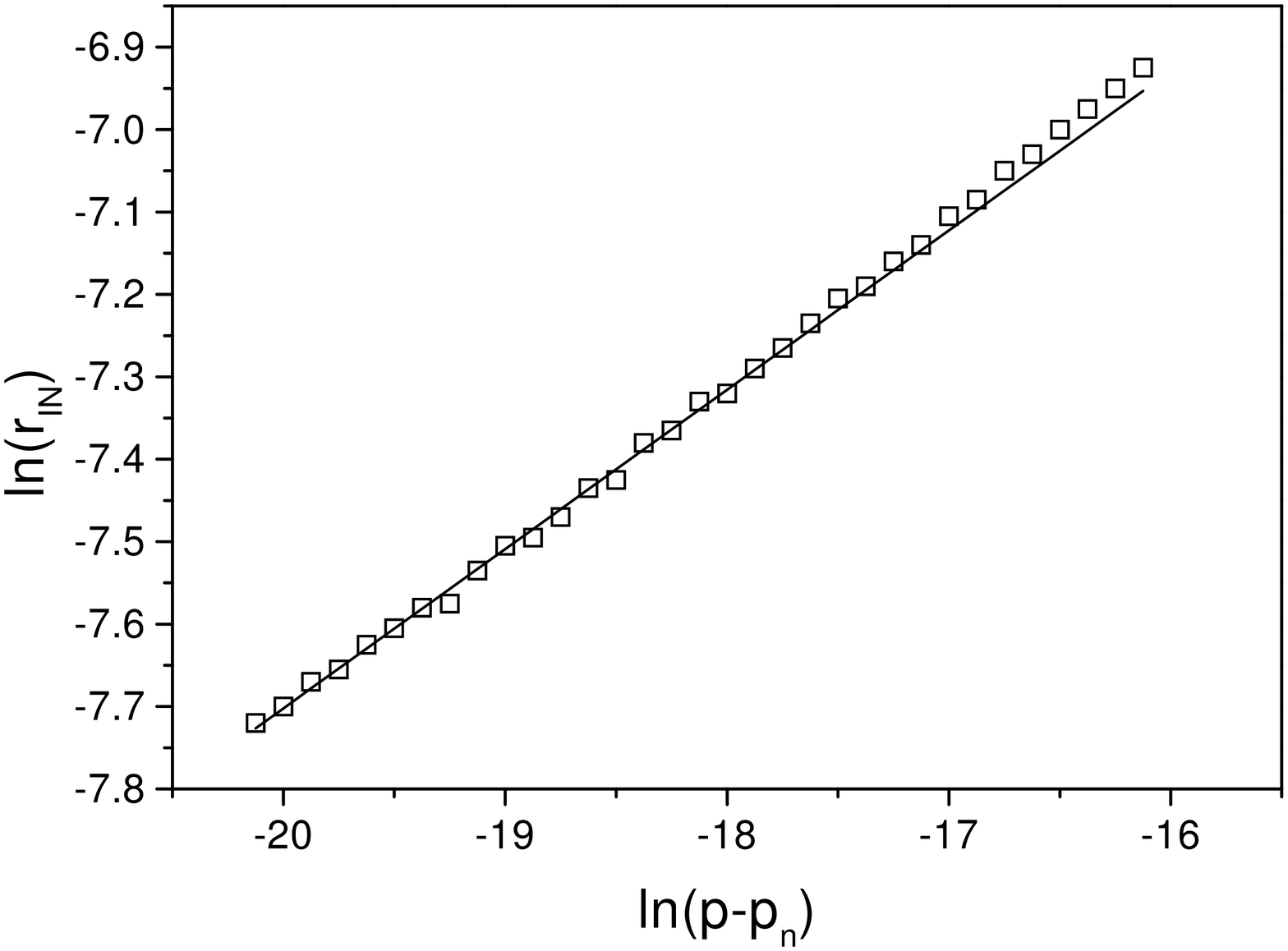}}
\centering \subfigure[]{
\includegraphics[width=0.34\textwidth,angle=-90]{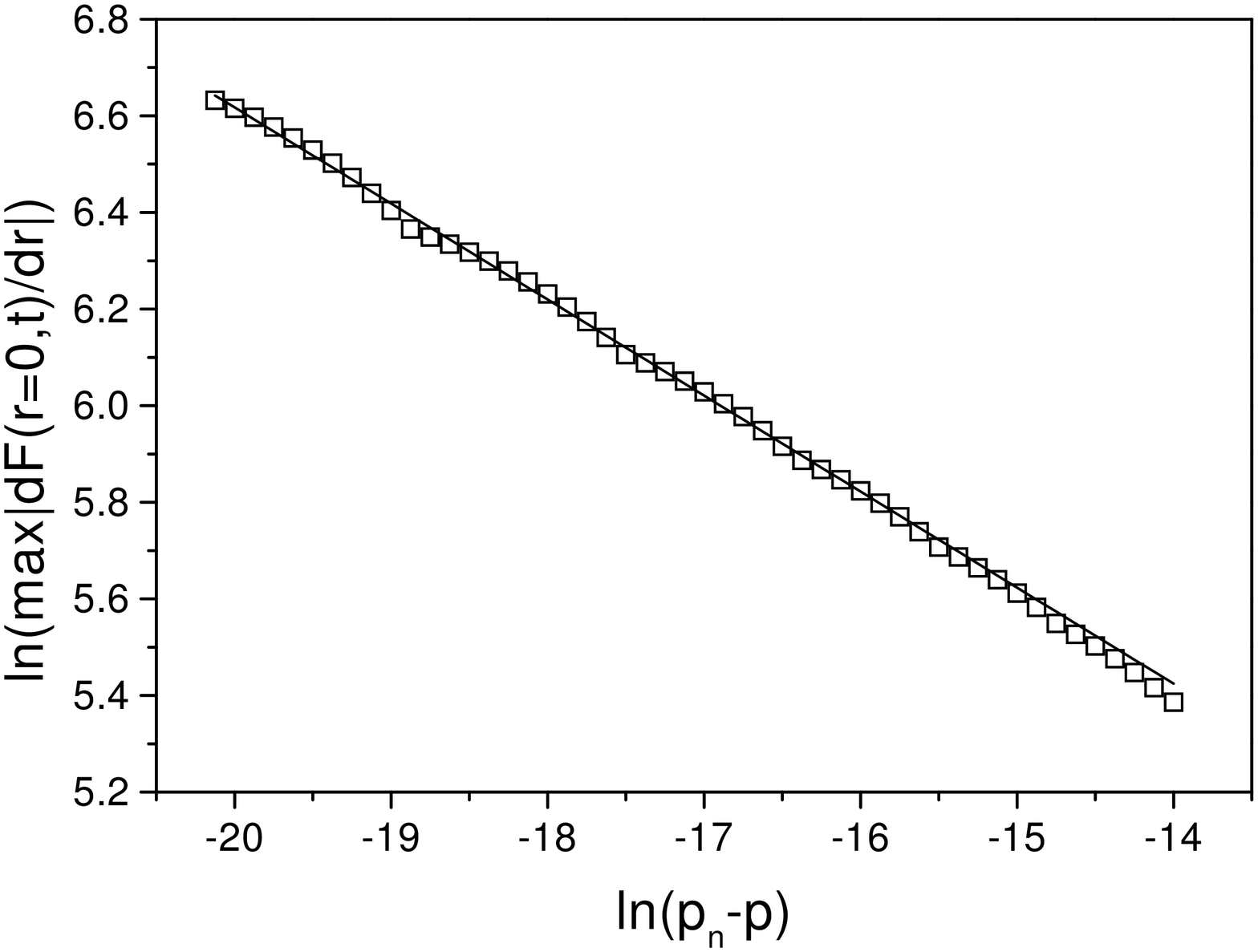}}
\caption{\small{$\eta=0.19$. (a) The supercritical (8) and (b)
subcritical (9) scalings around $p_n$ for $n=2$ (we get the same
picture for each $n$). The slopes of the linear fits are equal to
$\pm 0.195$ which agrees with the analytic prediction $\pm
1/\lambda_{CSS}$ where $\lambda_{CSS}(\eta=0.19)=5.1$ was obtained
independently from the linear perturbation theory by Lechner
\cite{ch}.} }\label{fig6}
\end{figure}

\subsection*{Interpretation of numerical results}
The results presented above confirm and extend the findings of
Lechner et al. \cite{w2}. Probing the bifurcation point $\eta_c$
with higher accuracy we improved the evidence that $\Delta$ diverges
as $\eta$ tends to the critical value $\eta_{c} = 0.17$ from above,
which in turn confirms the picture that the DSS cycle merges with
the CSS solution at the critical coupling constant $\eta_{c}$.
A natural question is:
  what is the meaning of the series of critical parameter values
$p_{n}$  within this picture?

We conjecture that our system shows a so called Shil'nikov
bifurcation \cite{k}. In his classification of  loop bifurcations
for three dimensional systems, Shil'nikov
 considered a system with a saddle  point together with
a homoclinic orbit which bifurcates for some value of a parameter.
Assuming that the eigenvalues of the saddle point are real and
satisfy the following conditions: $\lambda_{1}
> 0
> \lambda_{2}>\lambda_{3}$ and $\lambda_{1} + \lambda_{2}> 0$ (plus some less important technical conditions), Shil'nikov showed
that a saddle limit cycle bifurcates and the phase space picture
looks qualitatively as in Fig.~7a.

\begin{figure}[h]
\centering \subfigure[]{
\includegraphics[width=0.35\textwidth,angle=-90]{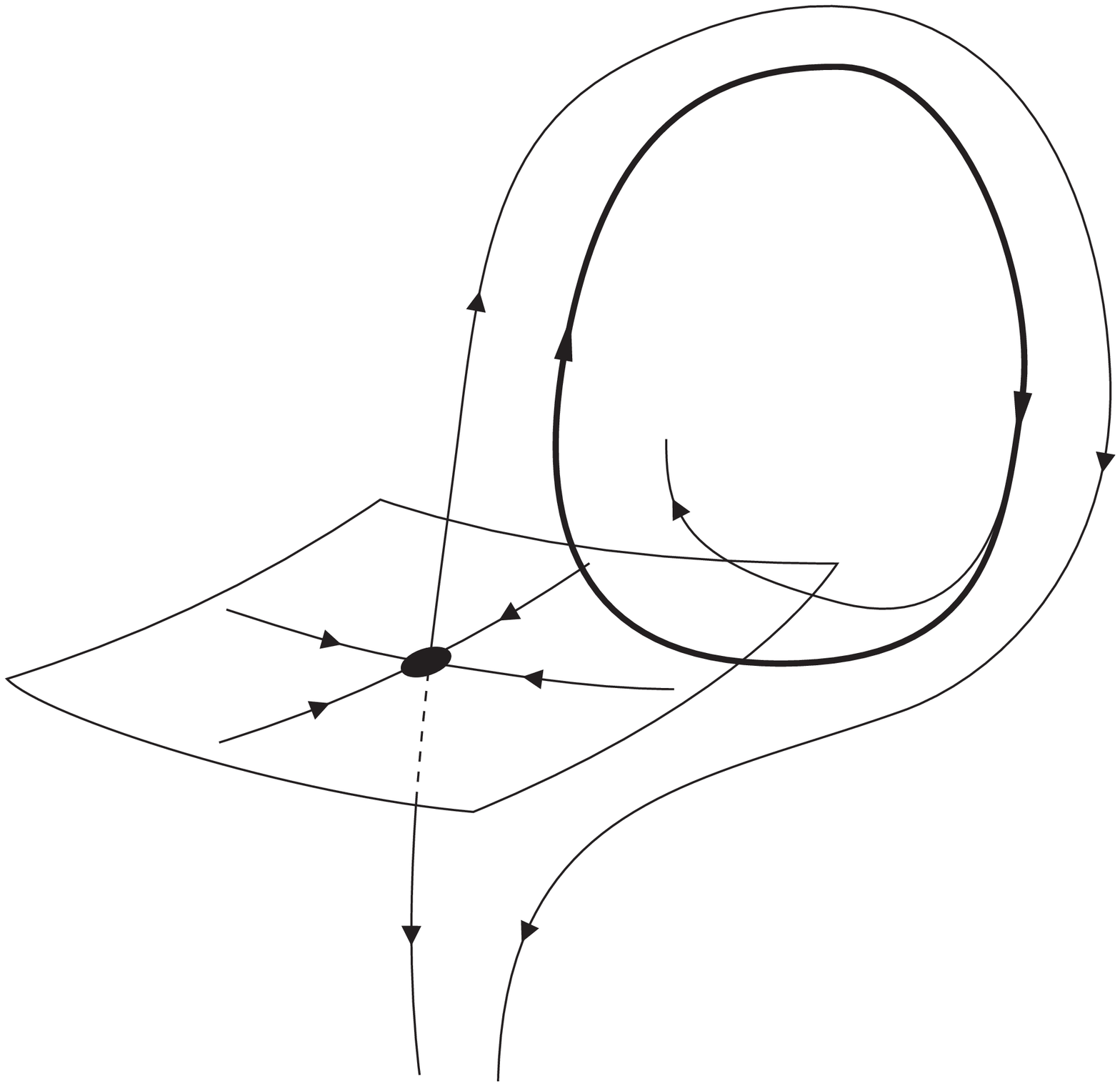}}
\subfigure[]{
\includegraphics[width=0.35\textwidth,angle=-90]{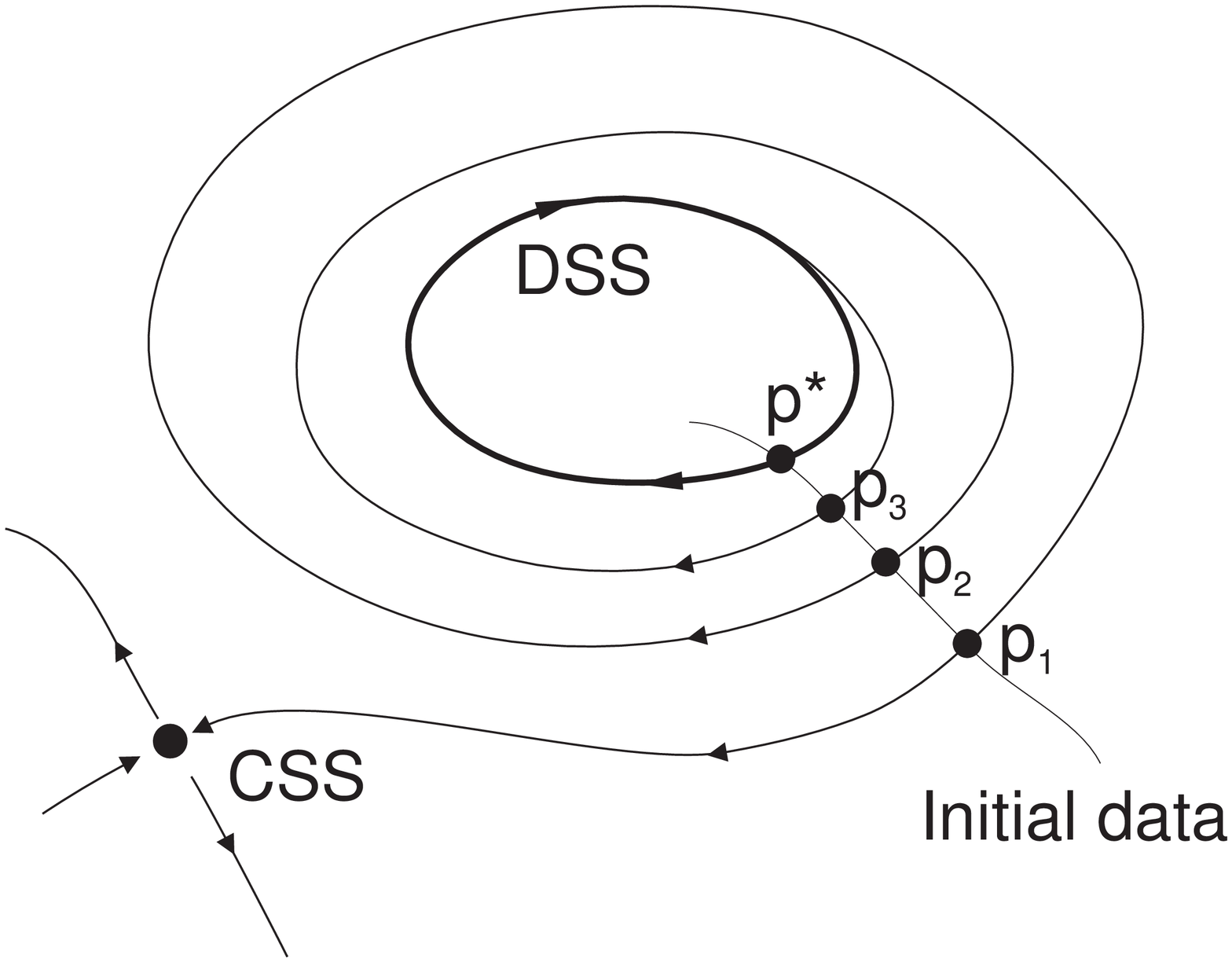}}
 \caption{\small{(a) Shil'nikov bifurcation. (b) The conjectured phase space picture.} }\label{fig6}
\end{figure}

Of course, our system is infinite dimensional and the Shil'nikov
theorem cannot be applied directly. Nevertheless, it is expected
that a similar picture to Fig.~7a will be valid for higher
dimensional systems as long as  only a few largest eigenvalues of
the perturbation matter. Recently, Donninger \cite{d} has studied
linear perturbations around the CSS solution and found that for
coupling constants around the critical value $\eta_c$ the first
three largest eigenvalues do in fact satisfy  the above stated
Shil'nikov conditions. Combining this property with the fact  that
the bifurcating DSS solution  is a saddle limit cycle, we conjecture
that the (one-dimensional) unstable manifold of the DSS solution
lies on the stable manifold of the neighboring CSS solution. This is
sketched in Fig.~7b.
More precisely, the DSS unstable manifold winds around the limit
cycle (infinitely many times) and eventually runs into the CSS
saddle. Suppose that a curve of initial data intersects this spiral
manifold at values $p_{n}$, with $\lim p_{n}= p^*$, where $p^*$
corresponds to the intersection with the limit cycle. Then, the
dynamical behavior will have exactly the form we observed above: for
$p$  equal to one of the $p_{n}$'s, the solution spirals $n$ times
around the limit cycle each time coming closer to the CSS--saddle
before hitting it. For $p = p_{n} \pm \varepsilon$, the behavior is
similar, except that the solution does not hit the CSS--saddle but
escapes along its unstable manifold. This is the reason why one
observes the CSS scaling around $p_n$ with an exponent related to
the unstable eigenvalue of the CSS solution. Note that the scaling
law (7) follows immediately from the picture shown in Fig.~7b
because during one cycle of evolution the distance from the DSS
limit cycle increases by the factor $e^{\lambda_{DSS} \Delta}$ (and
$\gamma_{DSS}=1/\lambda_{DSS}$).
\subsection*{Acknowledgments} This research was supported in part by
the Austrian Fonds zur F\"orderung der wissenschaftlichen Forschung
(FWF) Project P15738-PHY and in part by the Polish Ministry of
Science grant no. 1PO3B01229.

\end{document}